\documentclass[aps, prl
, twocolumn
, nofootinbib
, floatfix
, longbibliography
]{revtex4-1}
\usepackage{graphicx}
\usepackage{xcolor}
\usepackage[caption=false]{subfig}
\usepackage{mathrsfs,mathtools}
\usepackage{
amssymb}
\usepackage{physics}
\usepackage{bm}
\usepackage{braket}
\usepackage{listings}
\usepackage{cases}
\usepackage{here}
\usepackage{comment}
\usepackage{soul}
\usepackage{cancel}
\usepackage{cases}
\usepackage[utf8]{inputenc}
\usepackage{url}
\usepackage{longtable}
\usepackage[normalem]{ulem}
\usepackage{xspace}
\usepackage[colorlinks=true
,urlcolor=darkblue
,anchorcolor=darkblue
,citecolor=darkblue
,filecolor=darkblue
,linkcolor=darkblue
,menucolor=darkblue
,linktocpage=true
,pdfproducer=medialab
,pdfa=true
]{hyperref}

\newcommand{\Mpl}{M_\mathrm{Pl}}

\newcommand{\cs}{c_\mathrm{s}}

\newcommand{\GW}{\mathrm{GW}}

\newcommand{\uc}{\mathrm{c}}

\newcommand{\uf}{\mathrm{f}}

\newcommand{\calH}{\mathcal{H}}

\newcommand{\bfk}{\mathbf{k}}

\newcommand{\calN}{\mathcal{N}}
\newcommand{\calO}{\mathcal{O}}

\newcommand{\calP}{\mathcal{P}}

\newcommand{\ur}{\mathrm{r}}
\newcommand{\calS}{\mathcal{S}}

\newcommand{\bfx}{\mathbf{x}}

\newcommand{\bae}[1]{\begin{align} #1 \end{align}}
\newcommand{\bce}[1]{\begin{cases} #1 \end{cases}}
\newcommand{\dps}{\displaystyle}
\newcommand{\bfe}[4]{
\begin{figure} 
	\centering
	\includegraphics[#1]{#2}
	\caption{#3}
	\label{#4}
\end{figure}}

\definecolor{monza}{HTML}{CF000F}
\definecolor{darkblue}{HTML}{00008b}
\definecolor{darkmagenta}{HTML}{8b008b}

\newcommand{\eq}[1]{\begin{equation}\begin{split} #1 \end{split}\end{equation}}

\begin{document}
\title{Induced gravitational waves as a cosmological probe of the sound speed during the QCD phase transition}
\date{\today}

\author{Katsuya T. Abe}
\email{abe.katsuya@e.mbox.nagoya-u.ac.jp}
\author{Yuichiro Tada}
\email{tada.yuichiro@e.mbox.nagoya-u.ac.jp}
\author{Ikumi Ueda}
\email{ueda.ikumi@c.mbox.nagoya-u.ac.jp}
\affiliation{Department of Physics, Nagoya University, Nagoya 464-8602, Japan}

\begin{abstract}
The standard model of particle physics is known to be intriguingly successful. 
However their rich phenomena represented by the phase transitions (PTs) have not been completely understood yet, including the possibility of the existence of unknown dark sectors. 
In this Letter, we investigate the measurement of the equation of state parameter $w$ and the sound speed $\cs$ of the PT plasma with use of the gravitational waves (GWs) of the universe.
Though the propagation of GW is insensitive to $\cs$ in itself, the sound speed value affects the dynamics of primordial density (or scalar curvature) perturbations and the \emph{induced GW} by their horizon reentry can then be an indirect probe both $w$ and $\cs$.
We numerically reveal the concrete spectrum of the predicted induced GW with two simple examples of the scalar perturbation spectrum: the monochromatic and scale-invariant spectra.
In the monochromatic case, we see that the resonant amplification and cancellation scales of the induced GW depend on the $\cs$ values at different time respectively.
The scale-invariant case gives a more realistic spectrum and its specific shape will be compared with observations.
In particular, the QCD phase transition corresponds with the frequency range of the pulsar timing array (PTA) observations. If the amplitude of primordial scalar power is in the range of $10^{-4}\lesssim A_\zeta\lesssim10^{-2}$, the induced GW is consistent with current observational constraints and detectable in the future observation in Square Kilometer Array.
Futhermore the recent possible detection of stochastic GWs by NANOGrav 12.5 yr analysis~\cite{Arzoumanian:2020vkk} can be explained by the induced GW if $A_\zeta\sim\sqrt{7}\times10^{-3}$.
\end{abstract}

\maketitle

\section{Introduction}

The standard model of particle physics has so far achieved great success. It is, to a surprising extent, in agreement with many kinds of particle experiments.
On the other hand, it also theoretically reveals that particle physics is indeed filled with very rich and complicated phenomena which have not been well understood yet.
Several phase transitions in the standard model can be their representatives, and in particular, 
the quantum chromodynamics (QCD) phase transition is known as the most challenging physics due to its non-perturbative nature.
In Fig.~\ref{fig: g* and g*s}, we show a theoretical prediction of the effective degrees of freedom for energy density $g_*$ and entropy density $g_{*s}$ in thermal plasma at a given temperature $T$, defined by Eq.~\eqref{eq: g* and g*s}, with its theoretical uncertainty (cyan band)~\cite{Borsanyi:2016ksw,Saikawa:2018rcs}.
It shows that the uncertainty reaches $\sim10\%$ around 
$T\sim1\text{--}10\,\mathrm{GeV}$ even in full combination of analytic estimations and numerical lattice calculations (see Ref.~\cite{Saikawa:2018rcs} and references therein for detailed discussion).
It may be also possible that some unknown dark sector contributes to that phase transition. Their
further understanding requires more ``experimental" data.

Now the standard model of cosmology may be helpful as such an ``experiment". 
In an ordinary scenario, the universe is in fact thought to experience the history of phase transitions and thus can be a probe of these plasma's properties $g_*$ and $g_{*s}$, or equivalently the equation of state parameter (EoSp) $w=p/\rho$ and the sound speed (squared) $\cs^2=\partial p/\partial\rho$ with energy and pressure densities $\rho$ and $p$ (see Eq.~\eqref{eq: EoSp_cs2} for their transformation law), to which the cosmological perturbation has direct sensitivity.

\bfe{width=0.95\hsize}{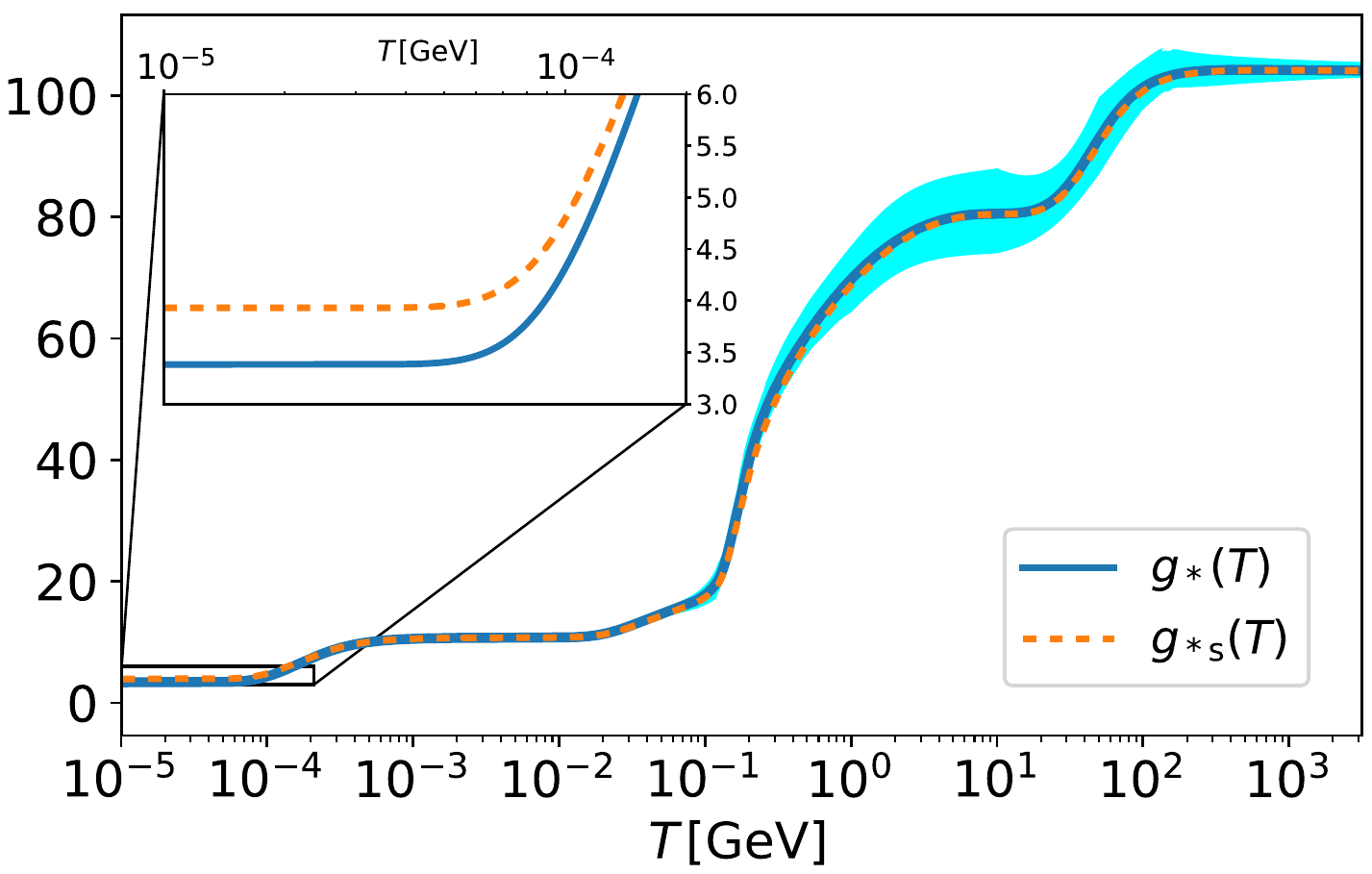}{Temperature dependence of the effective degrees of freedom for energy density $g_*$ (blue) and entropy density $g_{*s}$ (orange dotted) defined in Eq.~\eqref{eq: g* and g*s} and theoretical uncertainty in $g_*$ (cyan band), which are deeply investigated in Ref.~\cite{Saikawa:2018rcs}. We do not show the uncertainty in $g_{*s}$ as it cannot be distinguished from that in $g_*$ in this plot. Large uncertainty around $T\sim1\text{--}10\,\mathrm{GeV}$ is caused by the difficulty of QCD analysis. Three drops corresponding with the EW ($\sim100\,\mathrm{GeV}$), QCD phase transitions ($\sim0.1\,\mathrm{GeV}$), and the electron-positron annihilation ($\sim0.1\,\mathrm{MeV}$) change the fluid parameters $w$ and $\cs$ of the universe and affect the scalar (density) and tensor (GW) perturbation dynamics.
}{fig: g* and g*s}

The density (scalar) perturbation depends both on $w$ and $\cs$ (see its equation of motion~\eqref{eq: Bardeen}). However it cannot be a direct probe of them because the scalar perturbation has been diluted today by Silk damping on a relevant scale ($\lesssim1\,\mathrm{Mpc}^{-1}$)~\cite{Silk:1967kq}.
As a surviving probe, the stochastic gravitational waves (tensor perturbation) has been attracting more attentions after recent LIGO/Virgo collaboration's illustrious success of direct detection of gravitational waves (GWs)~\cite{Abbott:2016blz}.
The GW scientific community is now vigorously proceeding with
the ongoing/future plans of ground and space based GW detectors such as
LISA~\cite{2017arXiv170200786A}, 
Taiji/Tianqin~\cite{Guo:2018npi,Luo:2015ght}, 
DECIGO~\cite{Seto:2001qf,Yagi:2011wg}, 
AION/MAGIS~\cite{Badurina:2019hst}, 
KAGRA~\cite{Aasi:2013wya},
ET~\cite{EinsteinTelescope},
and pulsar timing arrays (PTAs)~\cite{Lentati:2015qwp,Shannon:2015ect,Arzoumanian:2018saf,Aggarwal:2018mgp,Arzoumanian:2020vkk,Moore:2014lga}.
In particular, the GW frequency $\sim10^{-8}\,\mathrm{Hz}$ corresponding with the Hubble scale during the QCD phase transition is in a sweetspot of PTA's sensitivity.
Recently NANOGrav collaboration announced that they have possibly succeeded to detect stochastic GWs for the first time in the PTA system~\cite{Arzoumanian:2020vkk}. It has been attracting attentions a lot (see, e.g., Refs.~\cite{Vaskonen:2020lbd,DeLuca:2020agl,Kohri:2020qqd,Bian:2020bps,Sugiyama:2020roc,Domenech:2020ers,bhattacharya2020implications} for a related GW source to our work).

Then how can be the stochastic GWs associated with phase transitions produced?
If the phase transition is first order, strong GWs can be produced via the bubble collision and so on
(see, e.g., the review~\cite{Cai:2017cbj} and references therein). However the phase transitions in the standard model are known to be crossover~\cite{Aoki:2006we,Kajantie:1996mn}.
The sufficient stochastic GWs may have been produced already during inflation from the quantum zero-point fluctuation similarly to the scalar perturbation.
Such GWs propagate as linear perturbations through phase transitions and can record the evolution of the EoSp $w$ in fact~\cite{Kuroyanagi:2008ye,Saikawa:2018rcs}.
They are however insensitive to the sound speed $\cs$. 

The recently refocused source of stochastic GWs is the second order effect of the scalar perturbation.
Though they are decoupled at linear order, the oscillation of scalar perturbations around/after their horizon reentry can generate tensor perturbations at second order~\cite{tomita1967non,Matarrese:1992rp,Matarrese:1993zf,Matarrese:1997ay,Carbone:2004iv,Ananda:2006af,Baumann:2007zm}.
Such \emph{scalar-induced GWs} can be a crosscheck of primordial black holes~\cite{Saito:2008jc,
Bugaev:2009zh, 
2010PThPh.123..867S, 
Bugaev:2010bb,
Inomata:2016rbd,
Bartolo:2018rku,
Bartolo:2018evs}
or a probe of the primordial scalar perturbation on a smaller scale~\cite{Inomata:2018epa}.
Their spectrum has been investigate not only on the pure radiational fluid but also on a more general cosmological background~\cite{Kohri:2018awv,Inomata:2019zqy,Inomata:2019ivs,Hajkarim:2019nbx,Domenech:2019quo,Domenech:2020kqm,Dalianis:2020cla,Domenech:2020ers}.
In particular, as the scalar perturbation depends on the sound speed, the induced GW can be an indirect probe both of the EoSp $w$ and the sound speed $\cs$.
In this Letter, we numerically calculate the spectrum of induced GWs affected by the QCD phase transition.

\section{Background evolution}\label{sec: bg_dynamics}

According to the standard Big-Bang cosmology, the universe experienced the high-energy radiation-dominated (RD) phase, where all the known particles are relativistic. As time goes, the radiation temperature however decreases and
some particles 
become non-relativistic, feeling their intrinsic masses or suddenly obtaining masses associated with 
symmetry breaking. In thermal equilibrium, non-relativistic particles 
soon disappear due to the Boltzmann suppression and cease to contribute to the radiation plasma. It is shown in terms of the effective degrees of freedom (DoF) $g_*$ and $g_{*s}$ in Fig.~\ref{fig: g* and g*s}. They are defined by the energy and entropy density $\rho$ and $s$ as
\bae{\label{eq: g* and g*s}
    \rho(T)=\frac{\pi^2}{30}g_*(T)T^4, \qquad s(T)=\frac{2\pi^2}{45}g_{*s}(T)T^3,
}
at temperature $T$. Fig.~\ref{fig: g* and g*s} shows their decreases three times corresponding to the electroweak (EW) phase transition ($\sim100\,\mathrm{GeV}$), the QCD phase transition ($\sim0.1\,\mathrm{GeV}$), and the electron-positoron annihilation ($\sim0.1\,\mathrm{MeV}$).

Such retreat of particles first affects the expansion law of the universe as a global dynamics. During the phase transition, non-relativistic particles still contribute to the energy density due to their rest masses before they completely disappear, while their pressure contributions corresponding to their momenta are getting suppressed. Therefore their ratio called the equation of state parameter (EoSp), $w=p/\rho$,
where $p$ is the pressure density, decreases from the pure radiational one $w=1/3$ and changes the dilution rate of the fluid energy through the continuity equation
\bae{\label{eq: continuity}
    \dv{\rho}{\eta}=-3(1+w)\calH\rho.
}
Here $\eta$ stands for the conformal time related with the cosmic time $t$ through the scale factor $a$ by $a\dd{\eta}=\dd{t}$. $\calH=a^\prime/a$ is the conformal Hubble parameter where the prime denotes the conformal time derivative. In addition to EoSp, the sound speed (squared) $\cs^2=\partial p/\partial\rho=p^\prime(T)/\rho^\prime(T)$, the other independent plasma parameter, can also affect the evolution of the scalar perturbation as we see in the next section.
These parameters are related with the effective DoF $g_*$ and $g_{*s}$ through
the first law of thermodynamics
\bae{\label{eq: first law of thermodynamics}
    p(T)=T s(T)-\rho(T),
}
as
\bae{
    \bce{
        \dps
        w(T) & 
        \dps\!\!
        = \frac{4g_{*s}(T)}{3g_*(T)}-1, \\[8pt]
        \dps
        \cs^2(T) &
        \dps\!\!
        = \frac{4\left(g_{*s}^\prime(T)T+4g_{*s}(T)\right)}{3\left(g_*^\prime(T)T+4g_*(T)\right)}-1.
    } \label{eq: EoSp_cs2}
}

Once the temperature dependence of $g_*$ and $g_{*s}$ is fixed, one can find the time evolution of the temperature of the universe by combining the continuity equation~\eqref{eq: continuity} with the Friedmann equation
\bae{
    \mathcal{H}^2 &= \frac{8\pi G}{3}a^2\rho, \label{eq: mathcal_H} 
} 
and the time evolution of the scale factor $a^\prime=a\calH$ by the definition of the Hubble parameter.\footnote{As all processes relevant to our work are reversible, the resultant solution for the scale factor should satisfy the entropy conservation $sa^3=\text{const.}$ until today. 
Inversely the scale factor evolution is often derived from this entropy conservation without directly solving the evolution equation.
However the phenomenological fitting formula for $g_*$ and $g_{*s}$ do not completely ensure the formal entropy conservation over the whole time. In this work, we then numerically solve the evolution equation also for the scale factor rather than imposing the entropy conservation.}
Making use of the lines shown in Fig.~\ref{fig: g* and g*s}
(they correspond to the fitting model derived in Ref.~\cite{Saikawa:2018rcs}), we show in Fig.~\ref{fig: EoS and cs2} the numerically obtained time evolution of $w$ and $\cs^2$.  
Though the uncertainty in $g_*$ and $g_{*s}$ would be inherited by $w$ and $\cs^2$, we only use these lines as a fiducial model.
They indeed show three drops from the radiational one $w=\cs^2=1/3$ corresponding to the EW and QCD transitions, and the electron-positron annihilation from left to right.
In the next section, we see their effect on the scalar and tensor perturbations and show how they are recorded onto the induced GWs. 
Note that the conformal time $\eta$ is normalized so that the current scale factor coincides with unity. It 
gives an implication of the comoving Hubble scale at each time, i.e., the considered perturbation with the comoving wavenumber $k$ enters the horizon when $k=\calH\simeq\eta^{-1}$.

We also mention the neutrino decoupling before closing this section. The standard model neutrinos cease to interact with photons around $T\sim1\,\mathrm{MeV}$ and the 
annihilation of electrons and positrons  
transfers their entropy only to photons after the neutrino decoupling. Hence the temperatures of photons and neutrinos start to deviate from each other. Thus one must separately take care of the fluid parameters for neutrinos and the other radiation in the first law of thermodynamics~\eqref{eq: first law of thermodynamics}, and then the pure radiational values $w(T)\simeq\cs^2(T)\simeq1/3$ can be consistently reproduced at low enough temperature $T\lesssim0.01\,\mathrm{MeV}$ even though the total $g_*$ and $g_{*s}$ do not converge to the same value as can be seen in the magnified panel of Fig.~\ref{fig: g* and g*s}.

\bfe{width=0.95\hsize}{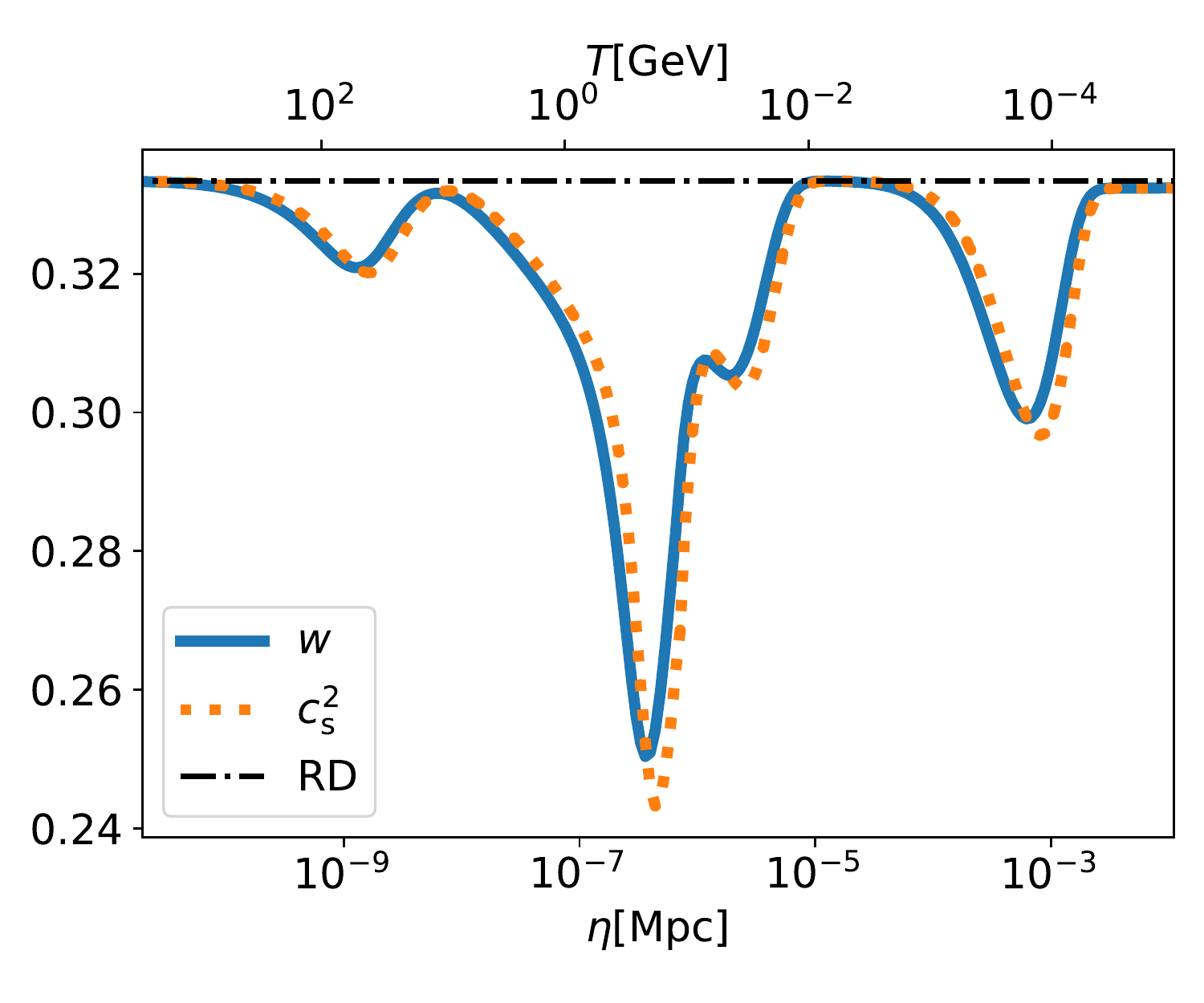}{The time evolution of the EoSp $w$ (blue) and the sound speed squared
$\cs^2$ (orange dotted). 
They correspond with the lines in Fig.~\ref{fig: g* and g*s} as a fiducial model.  
Three drops from the radiational one $w=\cs^2=1/3$ (black dot-dashed) correspond to the EW and QCD transitions, and the electron-positron annihilation from left to right, which 
affect the scalar and tensor perturbations.}{fig: EoS and cs2}

\section{Induced gravitational waves during the phase transition}

Let us then focus on perturbations. 
In the conformal Newtonian gauge,\footnote{As the induced GWs are second-order productions, it is known that there is a gauge dependence in their amplitude (see, e.g., Refs.~\cite{Tomikawa:2019tvi,DeLuca:2019ufz}). However such a difference disappears in the late time universe because the scalar perturbation itself is washed out by Silk damping and the tensor amplitude converges to the one calculated in the conformal Newtonian gauge~\cite{Inomata:2019yww}. Thus we can safely start with this gauge choice (see also Refs.~\cite{Ali:2020sfw,Chang:2020iji,Chang:2020mky} for recent discussions).}
the perturbed metric is defined by
\bae{
    \dd{s^2} &= a^2(\eta)\biggl\{-(1+2\hat{\Phi}(\eta,\bfx))\dd{\eta^2} \nonumber \\
    &\quad+\bqty{(1-2\hat{\Psi}(\eta,\bfx))\delta_{ ij}+\frac{1}{2}\hat{h}_{ij}(\eta,\bfx)}\dd{x^i}\dd{x^j}\biggr\}, \label{eq: metric}
}
where $\hat{\Phi}$ and $\hat{\Psi}$ are the scalar gravitational potential and curvature perturbation respectively,  
and $\hat{h}_{ij}$ is the traceless transverse tensor perturbation. The irrelevant vector perturbations are omitted. 
The hat indicates that they are originally quantum operators.
Below we study the tensor (GW) perturbations induced by the second-order correction of scalar perturbations, so that $h\sim\Phi^2\sim\Psi^2$ is expected for order counting.
Hereafter we assume that
the anisotropic stress is negligible during 
the RD era 
and we adopt $\hat{\Phi} = \hat{\Psi}$.\footnote{The anisotropic stress tensor is relevant around the neutrino decoupling and there the GW EoM~\eqref{eq: EoM_gw} should be modified~\cite{Kuroyanagi:2008ye,Saikawa:2018rcs}. In this work, we simply neglect their effect as the corresponding GW's scale ($k\sim10^4\,\mathrm{Mpc}^{-1}$ or $f\sim10^{-10}\,\mathrm{Hz}$) is 
marginally out of the region of interest with timing array 
PTA observations.}

In Fourier space, the perturbed Einstein equation at  
linear order in $h$ but quadratic order in $\Phi$ gives us the following
equation of motion (EoM) for the induced GWs: 
\bae{
    \Lambda_\eta\pqty{\!a(\eta)\hat{h}_\bfk(\eta)\!}\!:\!&=\!\bqty{\partial_\eta^2\!+\!k^2\!-\!\frac{1\!-\!3w(\eta)}{2}\calH^2(\eta)}\pqty{a(\eta)\hat{h}_{\bfk}(\eta)} \nonumber \\
    \!&=4a(\eta)\hat{\calS}_{\bfk}(\eta), \label{eq: EoM_gw}
}
where $\hat{\mathcal{S}}$ is a source term 
\eq{
\label{eq: source_term}
    &\hat{\mathcal{S}}_{\bfk}(\eta) = \int \frac{\dd[3]{\tilde{\bfk}}}{(2\pi)^3}e_{ij}(\bfk)\tilde{k}^i\tilde{k}^j\Biggl[2\hat{\Phi}_{\tilde{\bfk}}(\eta)\hat{\Phi}_{\bfk-\tilde{\bfk}}(\eta) \\
    &\quad\!+\!\frac{4}{3(1+w(\eta))}\pqty{\!\hat{\Phi}_{\tilde{\bfk}}(\eta)\!+\!\frac{\hat{\Phi}_{\tilde{\bfk}}^\prime(\eta)}{\mathcal{H}}\!}\!\pqty{\!\hat{\Phi}_{\bfk-\tilde{\bfk}}(\eta)  \!+\!\frac{\hat{\Phi}_{\bfk-\tilde{\bfk}}^\prime(\eta)}{\mathcal{H}}\!}\! \Biggl].
}
$e_{ij}(\bfk)$ is one polarization tensor.
At linear order (i.e. $\hat{\calS}\to0$), the GW EoM depends on the EoSp $w$ but not on the sound speed $\cs$. However the scalar perturbation is affected by both of them as can be seen in its EoM (see, e.g., Ref.~\cite{Hajkarim:2019nbx} for these equations):
\bae{
    \label{eq: Bardeen}
    &\hat{\Phi}_{\bfk}^{\prime\prime}(\eta)+3\mathcal{H}(1+\cs^2)\hat{\Phi}_{\bfk}^\prime(\eta)\nonumber \\
    &\qquad+\bqty{\cs^2k^2+3\mathcal{H}^2(\cs^2-w)}\hat{\Phi}_{\bfk}(\eta) = 0,
}
where we assume that there is no relevant entropic perturbation other than the adiabatic perturbation.
The induced GW thus can be an (indirect) probe not only of the EoSp $w$ but also of the sound speed $\cs$ of the plasma fluid.

We assume that the primordial tensor perturbations caused by the vacuum fluctuations are negligible. Then the above operator-level EoMs can be solved in the Green's function and the transfer function methods.
That is, the tensor perturbation is formally solved as
\eq{
\label{eq: h_k}
    \hat{h}_{\bm{\mathrm{k}}}&(\eta) = \frac{4}{a(\eta)}\int 
    \dd{\tilde{\eta}}
    G_k(\eta, \Tilde{\eta})\left[a(\Tilde{\eta})\hat{\mathcal{S}}_{\bm{\mathrm{k}}}(\Tilde{\eta})\right],
}
with the Green's function $G_\bfk(\eta,\tilde{\eta})$ associated with the GW EoM~\eqref{eq: EoM_gw}:
\begin{equation}\label{eq: green_func}
    \Lambda_\eta G_k(\eta,\tilde{\eta})=\delta(\eta-\tilde{\eta}).
\end{equation}
In a practical computation, such a Green's function can be obtained by the combination 
\begin{equation}\label{eq: method_gf}
    G_k(\eta, \Tilde{\eta}) =
    \frac{1}{\calN_k}\pqty{g_{1k}(\eta)g_{2k}(\Tilde{\eta})-g_{1k}(\Tilde{\eta})g_{2k}(\eta)}
\end{equation}
of two independent homogeneous solutions of Eq.~\eqref{eq: EoM_gw}: $\Lambda_\eta g_{1k}(\eta)=\Lambda_\eta g_{2k}(\eta)=0$.
The normalization $\calN_k=g_{1k}^\prime(\Tilde{\eta})g_{2k}(\Tilde{\eta})-g_{1k}(\Tilde{\eta})g_{2k}^\prime(\Tilde{\eta})$ is actually time independent, ensured by these homogeneous equations.
The scalar part is divided into the time-dependent transfer function $\Phi_k(\eta)$ and the primordial perturbation $\hat{\psi}_\bfk$ as $\hat{\Phi}_\bfk(\eta)=\Phi_k(\eta)\hat{\psi}_\bfk$.
$\hat{\psi}_\bfk$ is related with the gauge-invariant primordial curvature perturbation $\hat{\zeta}_\bfk$ by $\hat{\psi}_\bfk=-2\hat{\zeta}_\bfk/3$ at the sufficiently early RD universe and there the initial condition for the transfer function is given by $\Phi_k(\eta)\to1$ and $\Phi_k^\prime(\eta)\to0$ for $\eta\to0$.

Combining them, one obtains the power spectrum of the induced GW as
\bae{\label{eq: ps_vu}
    \calP_h(k,\eta)=\frac{64}{81a^2(\eta)}\int_{|k_1-k_2|\leq k\leq k_1+k_2}\hspace{-40pt}\dd{\log k_1}\dd{\log k_2}I^2(k,k_1,k_2,\eta) \nonumber \\
    \times\frac{\bqty{k_1^2-(k^2-k_2^2+k_1^2)^2/(4k^2)}^2}{k_1k_2k^2}\calP_\zeta(k_1)\calP_\zeta(k_2),
}
where
\bae{\label{eq: kernel_vu}
    &I(k,k_1,k_2,\eta)\!=\!k^2\int_0^\eta\!\dd{\tilde{\eta}}a(\tilde{\eta})G_k(\eta,\tilde{\eta})\biggl[2\Phi_{k_1}(\tilde{\eta})\Phi_{k_2}(\tilde{\eta}) \nonumber \\
    &\quad\left.+\frac{4}{3(1+w(\tilde{\eta}))}\pqty{\!\Phi_{k_1}(\tilde{\eta})\!+\!\frac{\Phi_{k_1}^\prime(\tilde{\eta})}{\calH(\tilde{\eta})}\!}\pqty{\!\Phi_{k_2}(\tilde{\eta})\!+\!\frac{\Phi_{k_2}^\prime(\tilde{\eta})}{\calH(\tilde{\eta})}\!}\!\right].
}
$\mathcal{P}_\zeta(k)$ is the power spectrum of the 
primordial curvature perturbation generated e.g. by 
inflation.
The dimensionless power spectrum is defined by
\bae{
    \braket{\hat{X}_\bfk\hat{X}_{\bfk^\prime}}=(2\pi)^3\delta^{(3)}(\bfk+\bfk^\prime)\frac{2\pi^2}{k^3}\calP_X(k),
}
for $X=h$ or $\zeta$.
The density parameter of GWs is given by the periodic average of this power spectrum $\overline{\calP_h(k,\eta)}$: 
\bae{
    \Omega_\GW(k,\eta)\coloneqq\frac{\rho_\GW(\eta,k)}{3\Mpl^2H^2(\eta)}=\frac{1}{24}\left(\frac{k}{aH}\right)^2\overline{\calP_h(k,\eta)},
}
as GWs periodically oscillate in time. 
Here $H=\calH/a$ is the ordinary Hubble parameter.

Practically the contribution to the kernel $I$~\eqref{eq: kernel_vu} is dominated around the time of scalars' horizon reentry. After the horizon cross, scalar perturbations rapidly decreases and
the GW density parameter 
almost settles down to a constant value
during the RD era because freely propagating GWs decay as $a^{-4}$ equally to
the radiation energy density. On the other hand, after the matter-radiation equality, the density parameter of induced GWs decreases as the universe expands because the matter energy density decreases ($\propto a^{-3}$) more slowly than the induced GWs. Moreover, even in the 
RD era, the slight deviation from the pure RD background due to the time dependence of $g_*$ and $g_{*s}$ affect the GW density parameter.
All these background evolutions can be included as the change of the Hubble parameter, and with the ($a^{-4}$)-dilution,
the current energy density of the induced GWs are given by
\begin{equation}
    \begin{aligned}\label{eq: omega_gw_k_eta0}
        \Omega_{\rm GW}(k, \eta_0)h^2 &=
        \pqty{\frac{a_\uc}{a_0}}^4\pqty{\frac{H_\uc}{H_0}}^2\Omega_{\rm GW}(k, \eta_{\rm c})h^2 \\
        &=\Omega_{\ur0}h^2\pqty{\frac{a_\uc\calH_\uc}{a_\uf\calH_\uf}}^2\frac{1}{24}\pqty{\frac{k}{\calH_\uc}}^2\overline{\calP_h(k,\eta_\uc)}.
    \end{aligned}
\end{equation}
Here we divided the computation process into two steps to avoid the time-consuming computation. The subscript ``$\uc$" represents the time well after the horizon reentry when the GW density parameter 
becomes almost constant (we adopt $k\eta_\uc=400$ throughout this work). Until this time we numerically solve the perturbation dynamics. The subscript ``$\uf$" indicates the time when $g_*$ and $g_{*s}$ are well reduced to the current values after all phase transitions but still in the RD era. Only the background dynamics is numerically solved until this time. After that until today labeled by the subscript ``$0$", the standard cosmology is assumed and simply gives the factor of the current radiation energy density parameter $\Omega_{\ur0}$.
We include the renormalized Hubble parameter of today $h=H_0/(100\,\mathrm{km}\,\mathrm{s}^{-1}\mathrm{Mpc}^{-1})$
as only the combination $\Omega_{\ur0}h^2=4.2\times10^{-5}$ can be directly determined by the current cosmic microwave background (CMB) temperature.

\begin{figure}
    \centering
    \includegraphics[width=0.95\hsize]{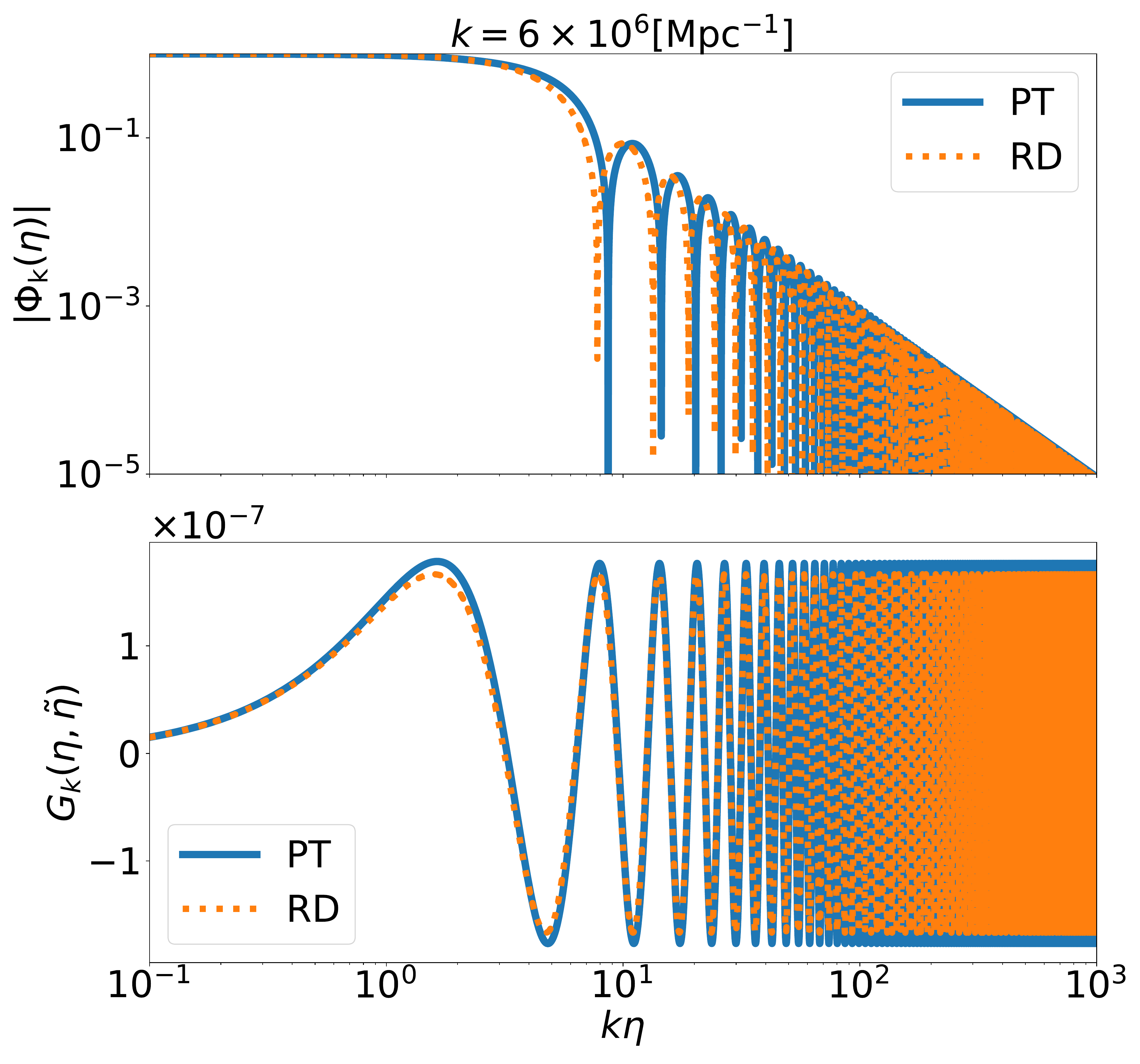}
    \caption{The numerical solutions (blue) for the scalar
    transfer function (upper panel) and the GW Green's function (lower panel) for $k=6\times10^6\,\mathrm{Mpc}^{-1}$ in the fiducial cosmological model (lines in Fig.~\ref{fig: g* and g*s}), as well as the pure RD solutions (orange dotted). 
    We use $\Tilde{\eta}\simeq1.67\times 10^{-9}\,\mathrm{Mpc}$ in the lower panel.}
    \label{fig: scalar_gf}
\end{figure}

We show in Fig.~\ref{fig: scalar_gf} the numerically obtained 
time evolution of the scalar transfer 
$\Phi_k(\eta)$ and the GW Green's function $G_k(\eta, \tilde{\eta})$
for $k = 6\times 10^6 \,\mathrm{Mpc^{-1}}$ corresponding to the QCD phase transition's scale as an example. 
The oscillation of scalar perturbation after its horizon reentry is slightly delayed from the pure radiational one because the sound horizon $\sim\cs/H$ becomes smaller during the QCD phase transition. Therefore  
particles can condense more until they feel pressure 
and larger GWs are induced. 
The GW Green's function is also slightly enhanced. 
That is because the homogeneous solution of Eq.~\eqref{eq: green_func} grows exponentially on a superhorizon scale as 
the effective frequency squared $k^2-(1-3w)\mathcal{H}^2/2$ can be negative for $w<1/3$.

We also mention a practical way to take the time average of the GW power spectrum. 
In principle, such a time average requires a further time integration of the kernel $I(k,k_1,k_2,\eta)$~\eqref{eq: kernel_vu} over one period around the evaluation time $\eta_\uc$.
However, with use of the mode-function expression~\eqref{eq: method_gf} of the Green's function, the kernel can be expanded as a mixing of two oscillating mode functions as
\bae{
    I(k,k_1,k_2,\eta)\!=\!I_2(k,k_1,k_2,\eta)g_{1k}(\eta)\!-\!I_1(k,k_1,k_2,\eta)g_{2k}(\eta),
}
with
\bae{
    &I_i(k,k_1,k_2,\eta)=\frac{k^2}{\calN_k}\int^\eta_0\dd{\tilde{\eta}}a(\tilde{\eta})
    g_{ik}(\tilde{\eta})\biggl[2\Phi_{k_1}(\tilde{\eta})\Phi_{k_2}(\tilde{\eta}) \nonumber \\
    &\quad+\!\frac{4}{3(1\!+\!w(\tilde{\eta}))}\pqty{\!\Phi_{k_1}(\tilde{\eta})\!+\!\frac{\Phi^\prime_{k_1}(\tilde{\eta})}{\calH(\tilde{\eta})}\!}\pqty{\!\Phi_{k_2}(\tilde{\eta})\!+\!\frac{\Phi^\prime_{k_2}(\tilde{\eta})}{\calH(\tilde{\eta})}\!}\biggr],
}
for $i=1$ and $2$. These coefficients $I_i$
depend on the evaluation time $\eta$ only through the integration upper limit.
As the scalar perturbation is damped enough well after the horizon cross, they are 
assumed to 
almost converge to constant values around the evaluation time $\eta_\uc$, 
and the integration kernel oscillates only by the mode functions $g_{1k}(\eta)$ and $g_{2k}(\eta)$.
Therefore its time average square can be approximated by
\bae{
    &\overline{I^2(k,k_1,k_2,\eta)}\simeq I_2^2(k,k_1,k_2,\eta)\overline{g_{1k}^2(\eta)} \nonumber \\
    &\qquad-2I_1(k,k_1,k_2,\eta)I_2(k,k_1,k_2,\eta)\overline{g_{1k}(\eta)g_{2k}(\eta)}\nonumber \\
    &\qquad+I_1^2(k,k_1,k_2,\eta)\overline{g_{2k}^2(\eta)},
}
where only the
time averages 
of the mode functions are needed. Hence one can much reduce the computational time in this way.

\subsection{Monochromatic case}\label{subsec: gw_spectrum_monochromatic}

\begin{figure}
    \centering
    \includegraphics[width=0.95\hsize]{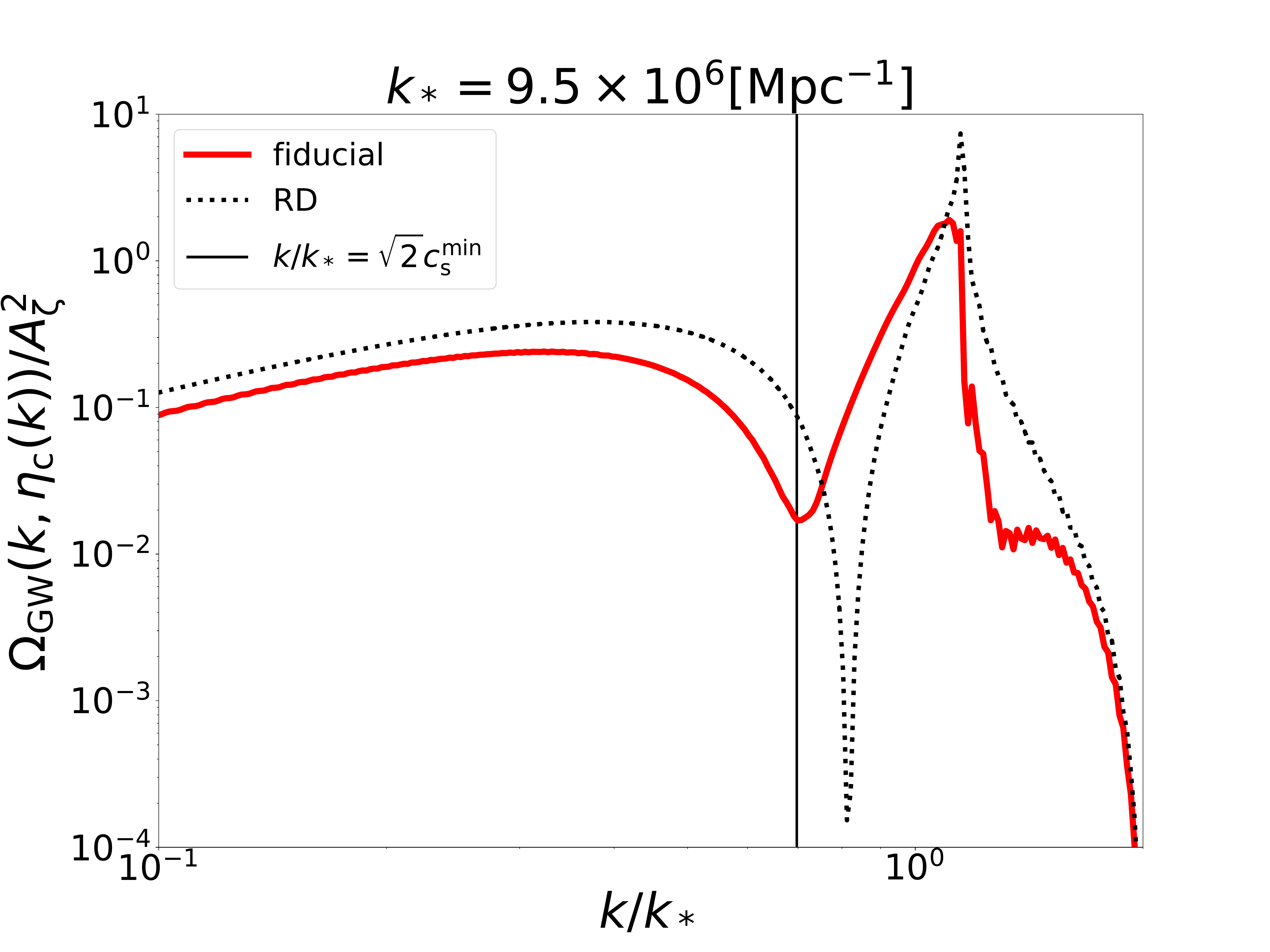}
    \caption{The GW density parameter 
    induced by the monochromatic scalar~\eqref{eq: monochromatic calPz} with $k_*=9.5\times10^6\,\mathrm{Mpc}^{-1}$ for the fiducial cosmological model (red)  
    as well as the pure RD case $w=\cs^2=1/3$ (black dotted). 
    They are normalized by the scalar amplitude square $A_\zeta^2$. 
    The evaluation time is $k\eta_{\rm c} = 400$ for each GW mode $k$. 
    The black
    vertical line indicates $\sqrt{2}\cs$  
    at $\eta_\mathrm{cancel}=4.48\times10^{-7}\,\mathrm{Mpc}$.  
    We checked that $k_*\eta_\mathrm{cancel}\sim4$ is 
    a universal relation of the cancellation scale for various $k_*$.
    On the other hand the peak scale 
    is not so different from the RD one because it can be considered to given by the average value of $\cs$ after the horizon reentry.}
    \label{fig: gw_mono_yokoyama}
\end{figure}

For more detailed understanding of the effect of 
$w$ and $\cs$, we first investigate  
the monochromatic scalar power spectrum, 
\bae{\label{eq: monochromatic calPz}
	\calP_\zeta(k)=A_\zeta\delta(\log k-\log k_*),
}
which picks up the single mode $k_*$. 
In Fig.~\ref{fig: gw_mono_yokoyama}, we show 
the resulting spectrum 
of induced GWs for 
the fiducial cosmological model 
as well as the pure RD universe $w=\cs^2=1/3$.
In the monochromatic case, the spectrum is known to exhibit one resonant amplification scale and one cancellation scale, which correspond to $k_\mathrm{peak}=2\cs k_*$ and $k_\mathrm{cancel}=\sqrt{2}\cs k_*$ respectively 
if $\cs$ is constant, reflecting the scalar's frequency $\cs k_*$~\cite{Domenech:2019quo}.
However, in more general cases where $\cs$ is time dependent,
these scales do not necessarily share the same value of $\cs$. 
For the cancellation scale, the (anti)resonance need only suppress the dominant contribution from the first oscillation of scalars. Thus $k_\mathrm{cancel}$ is determined by the $\cs$ value right after the scalar's horizon reentry (around $\eta_\mathrm{cancel}\sim4/k_*$).
On the other amplification side, the resonance should be kept well after the horizon reentry against the damping of scalars. The peak scale $k_\mathrm{peak}$ is hence controlled by the average value of $\cs$. These features can be seen well in Fig.~\ref{fig: gw_mono_yokoyama}.
We leave further investigations for future works.

\subsection{Scale-invariant case}\label{subsec: full_result}

\begin{figure}
    \centering
    \includegraphics[width=0.95\hsize]{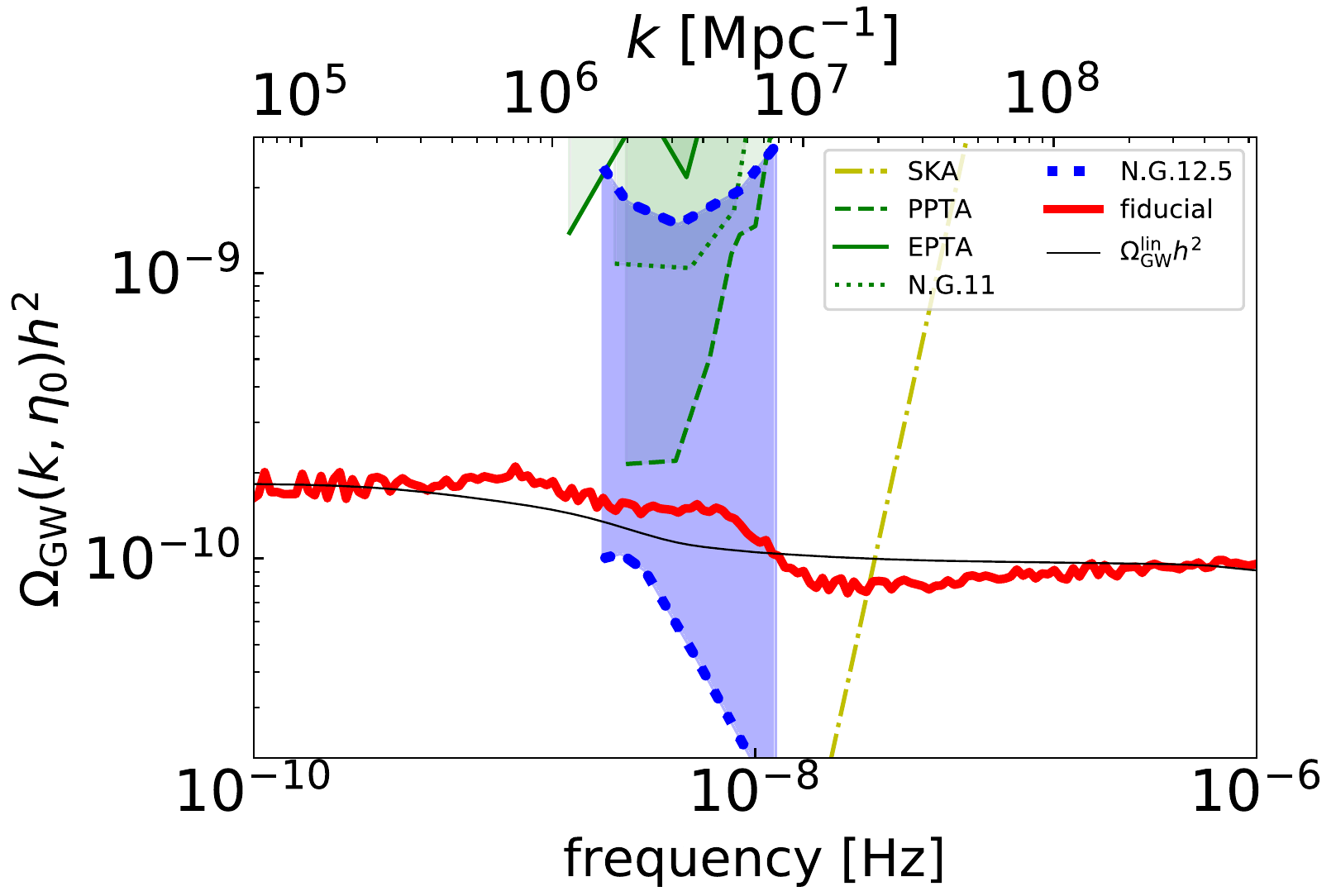}
    \caption{The spectral shape of the current energy density of induced GWs (red)  
    compared with that of the linear stochastic GW (black) studied in Refs.~\cite{Kuroyanagi:2008ye,Saikawa:2018rcs}. 
    The GW frequency $f$ is related with its wavenumber $k$ by $f=k/(2\pi)$. The noisy feature on the induced spectrum is merely caused by the numerical error. Here we assume $A_\zeta=\sqrt{7}\times10^{-3}$ for the induced GW and $A_h=18A_\zeta^2$ for the linear GW as an example.
    Green shaded regions show the observational exclusions by PPTA (dashed)~\cite{Shannon:2015ect}, EPTA (plane)~\cite{Lentati:2015qwp} and NANOGrav 11 yr (dotted)~\cite{Arzoumanian:2018saf,Aggarwal:2018mgp} respectively, and the yellow dot-dashed line indicates the prospect sensitivity by SKA~\cite{2009IEEEP..97.1482D,Moore:2014lga}.
    The blue-shaded region is the 2$\sigma$ consistency with the recent possible detection of the stochastic GW by NANOGrav 12.5 yr~\cite{Arzoumanian:2020vkk}.
    Our resultant induced GWs can be consistent with such detection.}
    \label{fig: omega_gw_full}
\end{figure}

Finally, we calculate the density spectrum of the induced GWs with the scale invariant power spectrum,
\bae{\label{def:scale_inv_ps}
    \mathcal{P}_{\zeta}(k) = A_{\zeta}.
}
Fig.~\ref{fig: omega_gw_full} shows the resulting current
spectrum of the induced GWs.
For a comparison we also show the spectral shape of the linearly-evolved GWs from scale-invariant primordial tensor perturbations $\calP^\mathrm{lin}_h(\eta\to0)\to A_h$ as investigated in Refs.~\cite{Kuroyanagi:2008ye,Saikawa:2018rcs}.
One first finds a common global step-like feature from left to right. It is caused by the reduction of $g_*$ and $g_{*s}$. As the effective degrees of freedom decreases, their entropy flows into the remained radiation, while the free propagating GW is unaffected. Hence the relative energy densities of GWs which entered the horizon earlier (i.e. higher frequency modes) get smaller than those of lower frequency modes.  
The linear GW is also affected by the EoSp $w$ on a superhorizon scale
in addition to this $g_*/g_{*s}$ effect as shown in the bottom panel of Fig.~\ref{fig: scalar_gf}.
On the other hand, the induced GW is 
affected both by $w$ and $\cs$. 
The resultant spectrum is then clearly distinguishable from that of the linear GW and can be a probe of the sound speed $\cs$ during the QCD phase transition. This is the main result of this work.

Note that the final GW density and hence its detectability in PTA depend on the scalar amplitude $A_\zeta$.
The current PTA constraint on the stochastic GW at $f\sim10^{-8}\,\mathrm{Hz}$ reads $\Omega_\GW h^2\lesssim10^{-9}$~\cite{Lentati:2015qwp,Shannon:2015ect,Aggarwal:2018mgp}.
On the other hand the
future Square Kilometer Array (SKA) project~\cite{2009IEEEP..97.1482D} expects the sensitivity improvement up to $\Omega_\GW h^2\gtrsim10^{-12}$.
Therefore if the scalar perturbation is somehow amplified as $10^{-4}\lesssim A_\zeta\lesssim10^{-2}$ on the QCD scale compared to $\calP_\zeta(k_\mathrm{CMB})\simeq2\times10^{-9}$ on the CMB scale $k_\mathrm{CMB}\sim0.05\,\mathrm{Mpc}^{-1}$~\cite{Aghanim:2018eyx},
the induced GW will be detectable.
In Fig.~\ref{fig: omega_gw_full}, we assume $A_\zeta=\sqrt{7}\times10^{-3}$ for induced GWs and $A_h=18A_\zeta^2$ for linear GWs as an example. Several current observational limits and future prospect are also shown for a comparison.
In particular, the recent possible detection of stochastic GWs by NANOGrav 12.5 yr is indicated in the blue band~\cite{Arzoumanian:2020vkk}.
Though it has not reached the sensitivity to resolve the detailed frequency dependence, it would be possible that the induced GW has been detected (see, e.g., Refs.~\cite{Vaskonen:2020lbd,DeLuca:2020agl,Kohri:2020qqd,Bian:2020bps,Sugiyama:2020roc,Domenech:2020ers,bhattacharya2020implications} for related discussions).

\section{Conclusion}

The stochastic GW induced by the scalar perturbations is investigated as a cosmological probe of the sound speed $\cs$ during the QCD phase transition.
Though the GW propagation itself does not depend on $\cs$, the induced GW can be an indirect probe of $\cs$ because the scalar perturbation is sensitive to $\cs$.
For illustration, we study two ideal spectra of the primordial scalar perturbation: monochromatic~\eqref{eq: monochromatic calPz} and scale-invariant ones~\eqref{def:scale_inv_ps}. 
Fig.~\ref{fig: gw_mono_yokoyama} shows the resultant spectral shape of the induced GW in the monochromatic case. There we found that the
cancellation scale is determined by the $\cs$ value right after the scalar's horizon reentry, $\eta_\mathrm{cancel}\sim4/k_*$,
while the amplification scale is controlled by the average value of $\cs$ after the horizon reentry because the resonance should be kept well against the damping of scalars.
Fig.~\ref{fig: omega_gw_full} is for the scale-invariant spectrum as the main result of this work.
As an indication of their $\cs$-dependence, the induced GWs (red) are enhanced at $\sim10^{-8}\,\mathrm{Hz}$ compared with the linearly evolved primordial GWs (black), corresponding to the $\cs$-reduction during the QCD phase transition.

The frequency range corresponding with the QCD phase transition is in a sweet spot of PTA GW observations.
If the scalar perturbations are somehow amplified to $10^{-4}\lesssim A_\zeta\lesssim10^{-2}$ on the corresponding scales compared to the CMB scale $\calP_\zeta(k_\mathrm{CMB}\sim0.05\Mpl^{-1})\simeq2\times10^{-9}$,
the amplitude of induced GWs will be at the detectable level with the future observation plan by SKA~\cite{2009IEEEP..97.1482D,Moore:2014lga} as well as consistent with the current observational upper bounds~\cite{Lentati:2015qwp,Shannon:2015ect,Arzoumanian:2018saf,Aggarwal:2018mgp}.
Recently NANOGrav collaboration announced their possible detection of stochastic GWs~\cite{Arzoumanian:2020vkk}, which could be explained by the induced GW if $A_\zeta\sim\sqrt{7}\times10^{-3}$.
Such amplification of scalar perturbations is motivated also from the viewpoint of primordial black hole (PBH).
The reduction of $w$ and $\cs^2$ during phase transitions makes the PBH formation easier (see Ref.~\cite{Byrnes:2018clq}). Furthermore the horizon scale during the QCD era corresponds with PBHs of $\calO(1)M_\odot$ mass, which can be another GW source via their mergers and detectable by ground-based observatories such as LIGO/Virgo collaboration~\cite{Bird:2016dcv,Clesse:2016vqa,Sasaki:2016jop,Sasaki:2018dmp}.
In these situations, further quantitative forecasts of $w$ and $\cs^2$, consistency observables amongst GW detector networks, etc., will be urgent tasks and we leave them for future works.

\acknowledgments

We are grateful to
Sachiko Kuroyanagi and 
Shuichiro Yokoyama
for helpful discussions.
This work is supported by JSPS KAKENHI Grants 
No. JP18J01992 (Y.T.), No. JP19K14707 (Y.T.),
and No. JP20J22260 (K.T.A).


\bibliography{main}

\end{document}